\documentclass[reprint,aps,prl,superscriptaddress]{revtex4-2}
\usepackage{graphicx}
\usepackage{dcolumn}
\usepackage{bm}
\usepackage[usenames]{color}
\usepackage{xcolor}
\usepackage{tabularx}
\usepackage{booktabs}
\usepackage{hyperref}
\usepackage{siunitx}
\usepackage[utf8]{inputenc}
\usepackage[english]{babel}
\hypersetup{colorlinks,linkcolor=red,citecolor=blue,urlcolor=blue,final}
\usepackage{amsmath, amsthm, amscd, amssymb}

\begin{document}



\title{Incoherent population trapping in quantum emitters}






\author{Sergei Lepeshov}
\affiliation{DTU Electro, Department of Electrical and Photonic Engineering, Technical University of Denmark, Ørsteds Plads 343, Lyngby, DK-2800, Kgs.\ Lyngby, Denmark}
\email{serle@dtu.dk}

\author{Søren Stobbe}
\affiliation{DTU Electro, Department of Electrical and Photonic Engineering, Technical University of Denmark, Ørsteds Plads 343, Lyngby, DK-2800, Kgs.\ Lyngby, Denmark}
\affiliation{NanoPhoton — Center for Nanophotonics, Technical University of Denmark, Ørsteds Plads 345A, Lyngby, DK-2800, Kgs.\ Lyngby, Denmark}
\email{ssto@dtu.dk}

\begin{abstract}
Deterministic emitters transform electronic excitations to photons with unity efficiency. Their development is crucial for both energy-efficient optical interconnects and photonic quantum technologies, but neither rigorous theoretical frameworks nor systematic experimental methods governing deterministic emitters and their identification were so far available. A central -- and seemingly obvious -- assumption underpinning previous works is that the radiative emission probability is proportional to the internal quantum efficiency. Here, we introduce a stochastic model of the decay dynamics in quantum emitters that disproves this assumption and provides a systematic framework for the development of deterministic quantum light sources. Our model agrees with a wide range of experimental findings, including time-resolved spectroscopy, autocorrelation measurements, and saturation spectroscopy, and it also explains a number of hitherto unexplained experiments. For example, our model shows that above-band continuous-wave excitation selects the exciton transitions with the lowest quantum efficiency, which is of direct importance for photonic quantum technologies relying on aligning nanostructures to emitters. We show that the underlying physics is governed by incoherent trapping of the population in metastable states, which has profound consequences for the physics of quantum emitters. Finally, we provide a straightforward experimental protocol for obtaining deterministic emitters.
\end{abstract}


\maketitle

The development of quantum light sources (QLSs) is central to quantum photonics, where they are enablers of both quantum communication~\cite{lu2021quantum,vajner2022quantum,yu2023telecom} and quantum computing~\cite{madsen2022quantum,geyer2024anisotropic,maring2024versatile}. A wealth of additional functionalities and components are enabled by using a QLS as a coherent light-matter interface, which, ideally, is also coupled to spin-based quantum memories~\cite{bhaskar2020experimental,chen2021scalable}. In addition, QLSs may also be operated at higher carrier densities or temperatures to enable nanolasers~\cite{jeong2020recent,dimopoulos2023experimental} or light-emitting diodes~\cite{shirasaki2013emergence,kim2024recent}. By placing a quantum emitter in a carefully engineered nanophotonic environment with a high Purcell factor, it is possible to realize QLSs with greatly improved emission rate, efficiency, coherence, brightness, and coupling efficiency to surrounding photonic circuitry~\cite{grange2017reducing,janitz2020cavity,antoniadis2023cavity,sapienza2015nanoscale}. This implies, in turn, that solid-state quantum emitters are particularly promising, and a wide range of emitters and materials are being explored, e.g., epitaxial quantum dots, colloidal quantum dots, vacancy defects in diamond, or implanted rare-earth emitters in silicon. Among these, epitaxial quantum dots (QDs) stand out for their high oscillator strength~\cite{albert2010quantum}, compatibility with semiconductor nanofabrication~\cite{ollivier2020reproducibility,garcia2021semiconductor}, and well-understood level structure comprised of bright and dark excitons, trions, biexcitons, and other excitonic complexes~\cite{kambhampati2011unraveling}. These are valuable resources of, e.g., single-photon pulses~\cite{lodahl2015interfacing} or entangled photon pairs~\cite{schimpf2021quantum,basso2021quantum}.

Previous works on QDs assumed that their brightness, which is a central figure of merit for a QLS as it quantifies the fraction of pulses containing one photon, is proportional to the internal quantum efficiency of the QD~\cite{senellart2017high}. Indeed, this is, by definition, also the case for an emitter in an excited state, which can be prepared by, e.g., a $\pi$-pulse applied to the QD in its ground state~\cite{zhai2020low}. However, the exciton can also decay non-radiatively, thus leaving an electron or a hole behind, or undergo a spin-flip into a dark-exciton state. While the decay processes leading to these metastable charge states are well-known and well-studied~\cite{erlingsson2001nucleus,khaetskii2000spin,tighineanu2013decay}, their impact on the excitation dynamics has been overlooked. These metastable charge states cannot decay radiatively before first undergoing another non-radiative decay back to either an empty QD or a bright exciton, but, crucially, the exciton can also not be excited by the same $\pi$-pulse as long as the QD is occupied by one of the metastable states due to Coulomb shifts~\cite{senellart2017high}.

\begin{figure*}[ht]
\includegraphics[width=1\linewidth]{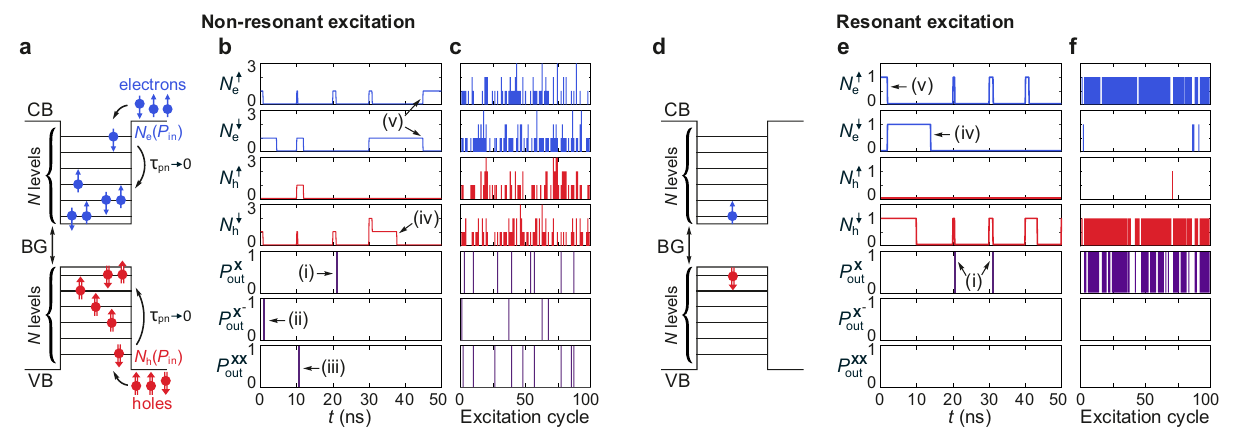} 
\caption{
Comparison of conventional and stochastic models of a semiconductor quantum dot (QD) under different excitation schemes. \textbf{a} Illustration of energy levels, their occupation by electrons and holes captured in the QD after above-band (non-resonant) excitation. \textbf{b} Time-evolution of the occupation of spin-up electrons ($N_\textrm{e}^\uparrow$), spin-down electrons ($N_\textrm{e}^\downarrow$), spin-up holes ($N_\textrm{h}^\Uparrow$), and spin-down holes ($N_\textrm{h}^\Downarrow$) as well as the emission from excitons ($P_{\textrm{out}}^\textbf{X}$), trions ($P_{\textrm{out}}^{\textbf{X}^-}$), and biexcitons ($P_{\textrm{out}}^\textbf{XX}$) during 5 cycles of the Monte-Carlo simulation. This exemplifies emission from various radiative excitonic states,
(i) the exciton ($\textbf{X}$), (ii) the negative trion ($\textbf{X}^{-}$), and (iii) the biexciton ($\textbf{XX}$) as well as (iv) non-radiative decay and (v) carrier spin flip. \textbf{c} Statistics of the population and the emission from exciton complexes under non-resonant excitation obtained over 100 cycles of Monte-Carlo evaluation. Notably, \textbf{b} and \textbf{c} are obtained for experimentally realistic parameters as explained in the main text and they indicate that despite the high quantum efficiency of $83\%$, the brightness is low.
\textbf{d}, 
\textbf{e}, and 
\textbf{f} Same as for \textbf{a}, 
\textbf{b}, and 
\textbf{c}
but for resonant $\pi$-pulse excitation of the exciton. Under these pumping conditions, the excitation of the exciton is much more efficient but non-radiative decays and spin flips still affect the QD. They lead to the population being trapped in dark states during which the excitation is off-resonance and show that the internal decay dynamics in a QD is a source of blinking, even under resonant excitation.
\label{Fig:1}
}
\end{figure*}

Although it is trivial to solve the rate equations governing the bright-dark exciton decay dynamics~\cite{tighineanu2013decay}, it is non-trivial to take higher-order quasi-particles and carrier population-dependent excitation into account, even though these effects can dramatically affect the QD brightness. This calls for a stochastic solution that treats the particle number deterministically at each moment of time while allowing for probabilistic evolution. Unlike the direct solution, which focuses solely on the decay of the exciton population, the stochastic solver effectively incorporates a photodetector in the modeling, which collapses the wavefunction of particles and allows deterministic treatment of the particle number, enabling calculations of the correlation functions and blinking.
Stochastic models have recently been providing new insights into the noise in nanolasers~\cite{bundgaard2023stochastic}. Here, we solve the coupled rate equations governing the decay dynamics using a Monte Carlo approach, and this leads to insights with serious ramifications, the most striking of which is that the brightness of a QD is not proportional to its internal quantum efficiency. A major technological consequence is that to build an ideal QLS by aligned construction of cavities around QDs, one must, surprisingly, not search for the brightest excitons as they will become comparatively dim in the presence of a sizable Purcell factor.

The starting point of our modeling of the carriers in a QD is the rate equations governing bright and dark excitons. This model does not include all metastable states (e.g., a single electron in a QD) and also not biexcitons and higher-order complexes, so we generalize it to include multiple electrons and holes distributed across $N$ energy levels. Figure~\ref{Fig:1}a shows the states in the conduction and valence bands along with an example of a higher-order state with multiple electrons and holes, each with either spin up ($\uparrow$, $\Uparrow$) or spin down ($\downarrow$, $\Downarrow$), where the single (double) arrow indicate the electron (hole) pseudospin~\cite{lodahl2015interfacing}. Complex excitonic states can be excited by above-band excitation, and they can undergo radiative recombination, non-radiative decay, and spin-flips with corresponding rates $\Gamma_{\textrm{r}}$, $\Gamma_{\textrm{nr}}$, and $\Gamma_{\textrm{sf}}$. All these rates are orders of magnitude slower~\cite{tighineanu2013decay} than the phonon-mediated intraband relaxation of carriers, which is therefore considered instantaneous. The Pauli principle is enforced by allowing only one electron or hole with a given energy and spin. It is challenging to calculate the Coulomb shifts and the various decay rates for excitonic complexes, but for our purposes, it is also not necessary. Instead, we assume that all charge configurations emit at different energies due to the Coulomb shifts and that the radiative, non-radiative, and spin-flip rates only depend on the population. These are reasonable assumptions for the experimental settings of QLSs: Low temperatures and populations on the order of or below saturation. As we will show, the agreement with experimental findings is excellent.

Generalizing the rate equations commonly studied for bright and dark excitons~\cite{lodahl2015interfacing} to a single-particle picture, we obtain,
\begin{equation}
\centering
\begin{split}
\frac{dN_\textrm{e}^{\uparrow}}{dt} = -N_\textrm{e}^{\uparrow}\Gamma_\textrm{nr}+\left(N_\textrm{e}^{\downarrow}-N_\textrm{e}^{\uparrow}\right)\Gamma_\textrm{sf}-\textrm{min}\left(N_\textrm{e}^{\uparrow},N_\textrm{h}^{\Downarrow}\right)\Gamma_\textrm{r}, \\
\frac{dN_\textrm{e}^{\downarrow}}{dt} = -N_\textrm{e}^{\downarrow}\Gamma_\textrm{nr}+\left(N_\textrm{e}^{\uparrow}-N_\textrm{e}^{\downarrow}\right)\Gamma_\textrm{sf}-\textrm{min}\left(N_\textrm{e}^{\downarrow},N_\textrm{h}^{\Uparrow}\right)\Gamma_\textrm{r}, \\
\frac{dN_\textrm{h}^{\Uparrow}}{dt} = -N_\textrm{h}^{\Uparrow}\Gamma_\textrm{nr}+\left(N_\textrm{h}^{\Downarrow}-N_\textrm{h}^{\Uparrow}\right)\Gamma_\textrm{sf}-\textrm{min}\left(N_\textrm{e}^{\downarrow},N_\textrm{h}^{\Uparrow}\right)\Gamma_\textrm{r}, \\
\frac{dN_\textrm{h}^{\Downarrow}}{dt} = -N_\textrm{h}^{\Downarrow}\Gamma_\textrm{nr}+\left(N_\textrm{h}^{\Uparrow}-N_\textrm{h}^{\Downarrow}\right)\Gamma_\textrm{sf}-\textrm{min}\left(N_\textrm{e}^{\uparrow},N_\textrm{h}^{\Downarrow}\right)\Gamma_\textrm{r},
\end{split}\label{eq1}
\end{equation}
where the subscripts $\textrm{e}$ and $\textrm{h}$ refer to electrons and holes, respectively. We consider pulsed excitation with a repetition rate of $T=10$~ns. For non-resonant excitation, the number of excited carriers, $N_{\textrm{e}}$ and $N_{\textrm{h}}$, follows Poissonian statistics with an expectation value $P_{\textrm{in}}$ referring to the input power and corresponding to the average number of generated electron-hole pairs. The coupled set of rate equations in Eq.~(\ref{eq1}) can be solved directly by Euler's method, but we apply instead a stochastic solver. Besides being conceptually simple, it brings two important advantages over the direct numerical solution: First, it provides insight into the time development of the populations and, therefore, allows identifying the quasiparticles that give rise to any particular emission event. Second, the average population can be extracted as an average over many excitation cycles, which provides direct and quantitative insight into the emission probability, in contrast to the -- often wrong -- assumption of a unity initial population required for the direct solution.

\begin{figure}
\centering\includegraphics[width=1\linewidth]{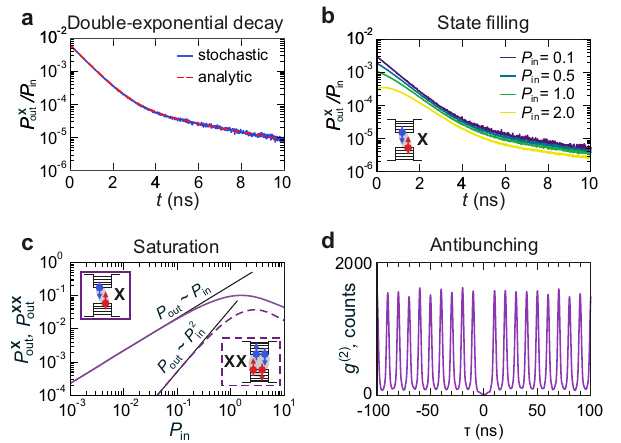} 
    \caption{Validation of the stochastic model.
 \textbf{a} Time-domain histogram of photon counts averaged over $10^9$ Monte Carlo cycles (blue solid) compared to the analytical solution for the bright-exciton emission~\cite{tighineanu2013decay} (red dashed) for an input power of $P_{\textrm{in}}=0.01$, which is well below saturation. \textbf{b} Time-domain histograms for different values of $P_\textrm{in}$ averaged over $10^8$ cycles exhibiting state filling upon increasing $P_\textrm{in}$. \textbf{c} Exciton (solid) and biexciton (dashed) emission probability as a function of  $P_\textrm{in}$, featuring exciton saturation at $P_\textrm{in}\approx 1.5$ and biexciton saturation at $P_\textrm{in}\approx 3$. \textbf{d} Second-order correlation function, $g^{(2)}$, calculated for $P_\textrm{in}=0.1$ showing antibunching of photon emission of the QD at $\tau=0$.
}
\label{Fig:2}
\end{figure}

We can now explore the interplay between excitation, radiative recombination, non-radiative decays, and spin-flips. We use the experimentally realistic parameters~\cite{wang2011mapping,chu2020lifetimes}
$\Gamma_{\textrm{r}} = 1$~ns$^{-1}$, $\Gamma_{\textrm{nr}} = 0.1$~ns$^{-1}$, $\Gamma_{\textrm{sf}} = 0.01$~ns$^{-1}$, and $P_{\textrm{in}}=1.5$, corresponding to an exciton quantum efficiency of $\eta_\text{QE,X} = \frac{\Gamma_{\textrm{r}}}{\Gamma_{\textrm{r}}+\Gamma_\textrm{nr}^\textbf{X}} \approx 83\%$, where $\Gamma_\textrm{nr}^\textbf{X}$ is the non-radiative decay rate of excitons, which is the quantity most commonly discussed in the QD literature. This is twice as fast as the corresponding single-particle rates due to the two-particle nature of the exciton~\cite{tighineanu2013decay}, i.e., $\Gamma_\textrm{nr}^\textbf{X} = 2\Gamma_\textrm{nr}$. A similar relation holds for the spin-flip rate. For these parameters, we obtain the population time series shown in Fig.~\ref{Fig:1}b. Evidently, the various excitonic states, such as excitons ($X$), negative trions ($X^-$), and biexcitons ($XX$), are only rarely populated. Crucially, the probability of emission from any one of the excitonic quasiparticles
appears to be low. This is confirmed by Fig.~\ref{Fig:1}c, which shows the population over 100 excitation cycles: Despite the high quantum efficiency, the exciton only emits eight photons during the 100 cycles. This is conceptually unsurprising because, by definition, the exciton is not present when the QD is occupied by any of the other charge states, but the magnitude is nevertheless surprising. We will return to precise calculations of $P_\textrm{out}^\textbf{X}$, but we can already conclude that a QD under above-band excitation is not a deterministic QLS.

When turning to resonant excitation, we assume that the pump is resonant with and polarized to excite only one of the bright exciton states. In this case, an empty QD is excited with a probability of unity by a $\pi$-pulse as shown in Fig.~\ref{Fig:1}d. Under this excitation condition, higher-order excitonic complexes cannot be excited, so this has previously been assumed to turn the QD into a deterministic QLS. However, our calculations of the population time series displayed in Fig.~\ref{Fig:1}e show that the resonantly excited exciton can still undergo non-radiative decays and spin-flips, which prevents further resonant excitation until the QD is empty. This significantly impacts the probability of emission over many excitation cycles, as shown in Fig.~\ref{Fig:1}f. Although brighter than a non-resonantly excited QD, this shows that a QD, even under resonant excitation, is not a deterministic QLS.

Before exploring the wider impact of our model, we first validate it against known theory and established experimental findings. Running the model across $10^9$ cycles allows us to obtain a decay curve of $P_\textrm{out}^\textbf{X}$ corresponding to the exciton emission commonly detected in time-correlated experiments on QDs for an integration time of $10$~s. Figure~\ref{Fig:2}a shows a photoluminescence-intensity decay curve in the non-resonant excitation scheme normalized by $P_\textrm{in}=0.01$, exhibiting the double-exponential decay characteristic of excitons~\cite{lodahl2015interfacing}. The decay curve is in excellent agreement with the analytical solution to the bright-exciton population given by the well-known theoretical model for bright and dark excitons~\cite{lodahl2015interfacing,tighineanu2013decay}. Tracking the carrier population in the stochastic simulations show that the fast exponent with a decay rate of $\Gamma_\textrm{f}\simeq\Gamma_\textrm{r}+2\Gamma_\textrm{nr}$ is due to the bright exciton recombination while the slow exponent with $\Gamma_\textrm{s}\simeq2\Gamma_\textrm{nr}+2\Gamma_\textrm{sf}$ is due to the dark exciton, which agrees with well-known results~\cite{lodahl2015interfacing}.  

\begin{figure*}
\centering\includegraphics[width=1.0\linewidth]{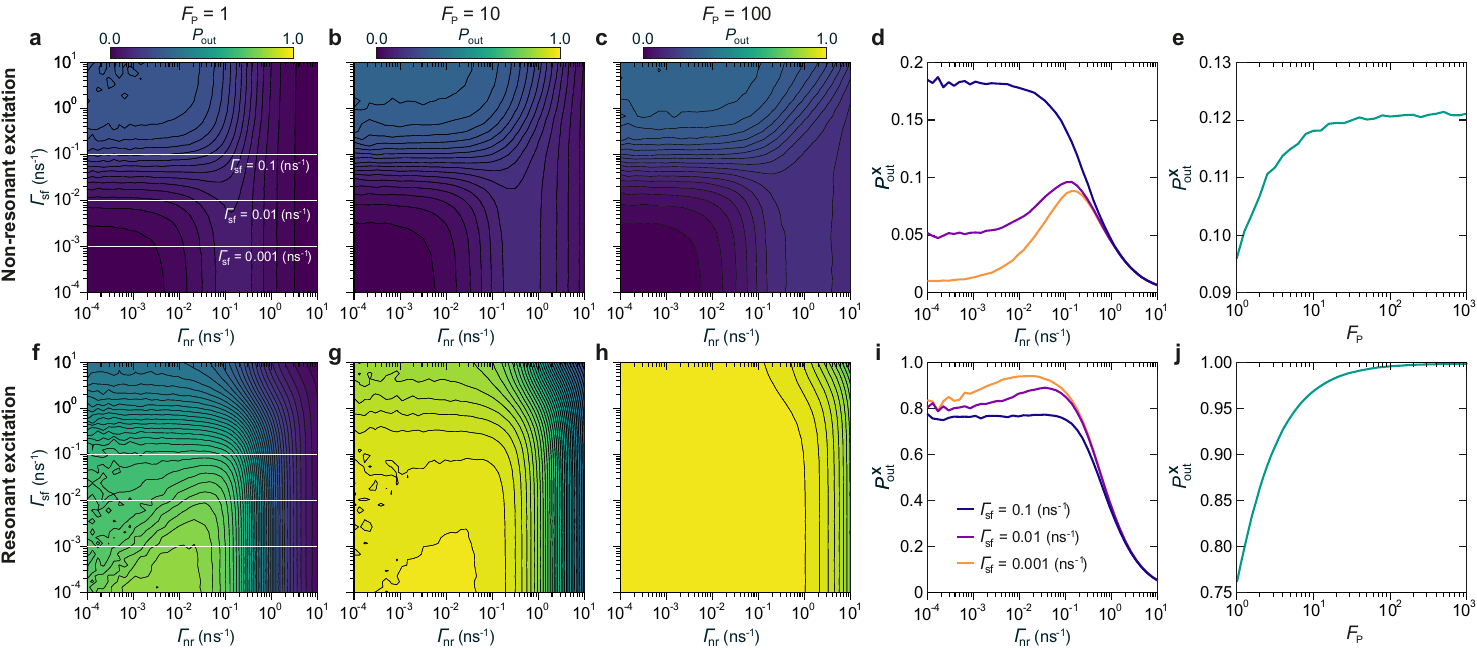} 
    \caption{Probability of emission from excitons for non-resonant (top panels) and resonant excitation (bottom panels). \textbf{a}-\textbf{c} Probability of emission, $P_\text{out}^X$ as a function of $\Gamma_\textrm{nr}$ and $\Gamma_\textrm{sf}$ for three different Purcell factors, $F_\text{P}$. \textbf{d} Probability of emission for $F_\text{P}=1$ as a function of $\Gamma_\textrm{nr}$ for three values of $\Gamma_\textrm{sf}$ corresponding to the white lines in \textbf{a}. For the spin-flip rates commonly observed in experiments, i.e., $\Gamma_\textrm{sf} < 0.1 \text{ns}^{-1}$, the emission probability is maximum exactly around the commonly observed $\Gamma_\textrm{nr}$.
    \textbf{e} Emission probability as a function of the Purcell factor. \textbf{f}-\textbf{j} Same plots as \textbf{a}-\textbf{e} but for resonant excitation, where we observe a maximum in the emission probability when $\Gamma_\textrm{sf}<\Gamma_\textrm{nr}<\Gamma_\textrm{r}$, i.e., a lower non-radiative decay rate does not always imply a brighter QD. Furthermore, the values of $\Gamma_\textrm{nr}$ and $\Gamma_\textrm{sf}$ for which the emission probability is maximized are different for non-resonant and resonant excitation. All presented data are obtained with $\Gamma_\text{r}=1~\text{ns}^{-1}\times F_\textrm{P}$ and a repetition duration of $T = 10~\text{ns}$. In panels \textbf{e} and \textbf{j}, $\Gamma_\textrm{nr}=0.1~\text{ns}^{-1}$ and $\Gamma_\textrm{sf}=0.01~\text{ns}^{-1}$.}
\label{Fig:4}
\end{figure*}

\begin{figure}\centering\includegraphics[width=1.0\linewidth]{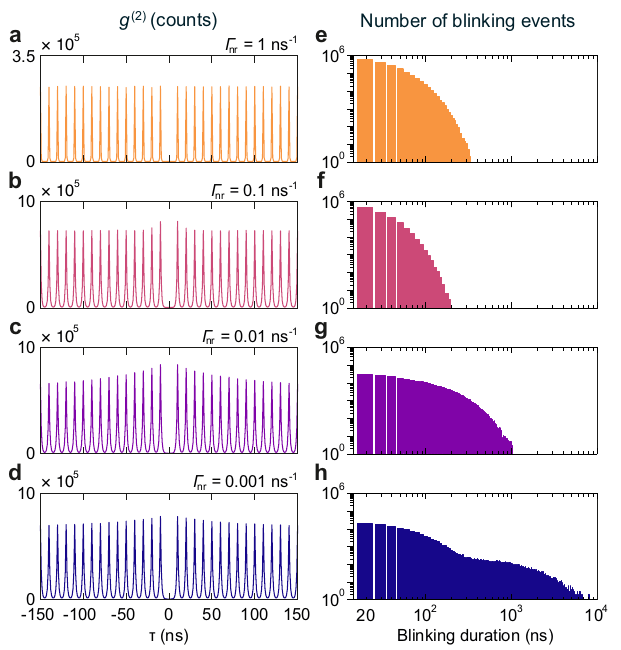} 
    \caption{
Second-order correlation function and blinking duration statistics in the resonant excitation regime. \textbf{a}-\textbf{d} Second-order correlation function for different values of $\Gamma_\textrm{nr}$: \textbf{a} $\Gamma_\textrm{nr} = 1~\textrm{ns}^{-1}$, \textbf{b} $\Gamma_\textrm{nr} = 0.1~\textrm{ns}^{-1}$, \textbf{c} $\Gamma_\textrm{nr} = 0.01~\textrm{ns}^{-1}$, and \textbf{d} $\Gamma_\textrm{nr} = 0.001~\textrm{ns}^{-1}$. \textbf{e}-\textbf{h} Histograms of blinking duration as a function of $\Gamma_\textrm{nr}$ same as for \textbf{a}-\textbf{d}. The presented data are acquired during $10^7$ excitation cycles.}
\label{Fig:3}
\end{figure}

Figure~\ref{Fig:2}b shows the decay curves of $P_\textrm{out}^\textbf{X}$ for different values of $P_\textrm{in}$, which are double-exponential and with pronounced filling for higher input powers and shorter times. Notably, the impact of filling on the decay curves is not captured by the analytical model, as it incorrectly assumes an excited exciton as the initial condition. This filling effect is due to saturation of the quantum states, which is confirmed by Fig.~\ref{Fig:3}c, showing  $P_\textrm{out}^\textbf{X}$ and $P_\textrm{out}^\textbf{XX}$ as functions of $P_\textrm{in}$.
As expected, $P_\textrm{out}^\textbf{X}$ scales linearly with power for $P_\textrm{in}<10^{-1}$, while $P_\textrm{out}^\textbf{XX}$ scales as $P_\textrm{in}^2$ for $P_\textrm{in}<0.5$, which reflects the well-known single- and bi-molecular nature of excitons and biexcitons, respectively~\cite{simeonov2008complex}.
Moreover, we show that our model can be used to calculate the second-order correlation function,$g^{(2)}$, of the light emitted from a QD, which we define as $g^{(2)}=\langle n_{\delta t}^\textrm{(i)}(t) n_{\delta t}^\textrm{(ii)}(t+\tau) \rangle$, where $n_{\delta t}$ is the number of photons observed at one of the photodetectors, either (i) or (ii), within a time interval, $\delta t=1$~ns, and $\tau$ is the time delay. Figure~\ref{Fig:2}d shows $g^{(2)}$ for the exciton emission under non-resonant excitation and at $P_\textrm{in}=0.1$, which is well below the saturation threshold at $P_\textrm{in}=1.5$. We observe pronounced anti-bunching at $\tau=0$ characteristic of single-photon quantum statistics.
These results show that our stochastic model, unlike other models employing rate equations, can describe not only steady-state dynamics but also filling, saturation, different pumping schemes, and quantum statistics.

After validating our model, we proceed to explore the probability of emission from one of the bright excitons, $P_\textrm{out}^\textbf{X}$, which we define as the average number of photons emitted from the exciton after averaging over $10^6$ excitation cycles. For a QD prepared in the bright-exciton state, it is generally assumed~\cite{senellart2017high} that $P_\textrm{out}^\textbf{X} = \eta_\text{QE,X}$, but we find that this exciton state can only be prepared deterministically in a few limiting cases. Figure~\ref{Fig:4} shows the probability of emission for non-resonant excitation (Fig.~\ref{Fig:4}a-e) and resonant excitation (Fig.~\ref{Fig:4}f-j) for a wide range of non-radiative, spin-flip, and radiative decay rates, where the latter is modeled as a Purcell factor multiplied only on the radiative decay rate of the exciton relative to the bulk value of $\Gamma_\text{r} = 1\: \text{ns}^{-1}$ while leaving all other radiative decay rates at the bulk value.

For non-resonant excitation, we observe a maximum in $P_\textrm{out}^\textbf{X}$ around $\Gamma_\textrm{nr} \simeq 0.2\: \text{ns}^{-1}$ regardless of $\Gamma_\textrm{sf}$ as long as 
$\Gamma_\textrm{sf} < 0.03 \text{ns}^{-1}$. This effect is strongly dependent on the Purcell factor (Figs.~\ref{Fig:4}a-c) for experimentally relevant spin-flip rates but vanishes for$\Gamma_\textrm{sf} > 10^{-1}\text{ns}^{-1} $ as shown in Fig.~\ref{Fig:4}d. This implies that time-resolved spectroscopy on QD ensembles would predominantly measure $\Gamma_\textrm{nr} \simeq 0.2\: \text{ns}^{-1}$ even with considerable variation in $\Gamma_\textrm{nr}$ across the ensemble because excitons with this non-radiative decay rate would be brightest. As we will show later, this prediction is in excellent agreement with experiments.
Figure~\ref{Fig:4}e quantifies what we have already concluded qualitatively before, namely that a QD under non-resonant excitation is far from a deterministic QLS, but it adds the surprising insight that the Purcell factor can only provide minor improvements to the emission probability in this regime. The reason is that, for higher Purcell factors, the non-radiative decay rate should be higher to reach maximum brightness.

Figure~\ref{Fig:4}f-j show $P_\textrm{out}^\textbf{X}$ calculated for resonant excitation, which is more relevant for applications in quantum technologies. We find that the maximum of $P_\textrm{out}^\textbf{X}$ appears for very different values of
$\Gamma_\textrm{nr}$ and 
$\Gamma_\textrm{sf}$ as compared to the non-resonant case. For a Purcell factor of $F_\text{P}=1$ (Fig.~\ref{Fig:4}f), a smaller $\Gamma_\textrm{nr}$ only implies a higher emission probability if $\Gamma_\textrm{r}>\Gamma_\textrm{nr}>\Gamma_\textrm{sf}$. This exemplifies again that a high internal quantum efficiency does not imply a high emission probability and, therefore, also not bright QLSs, see Fig.~\ref{Fig:4}i. Contrary to the case of non-resonant excitation, a higher Purcell factor (Fig.~\ref{Fig:4}g,h,j) can turn the QD into a deterministic QLS, but we note that this requires rather high Purcell factors, e.g., $P_\textrm{out}^\textbf{X}$ exceeds $99\%$ only for $F_\text{P}>30$.

\begin{figure}
    \centering\includegraphics[width=1.0\linewidth]{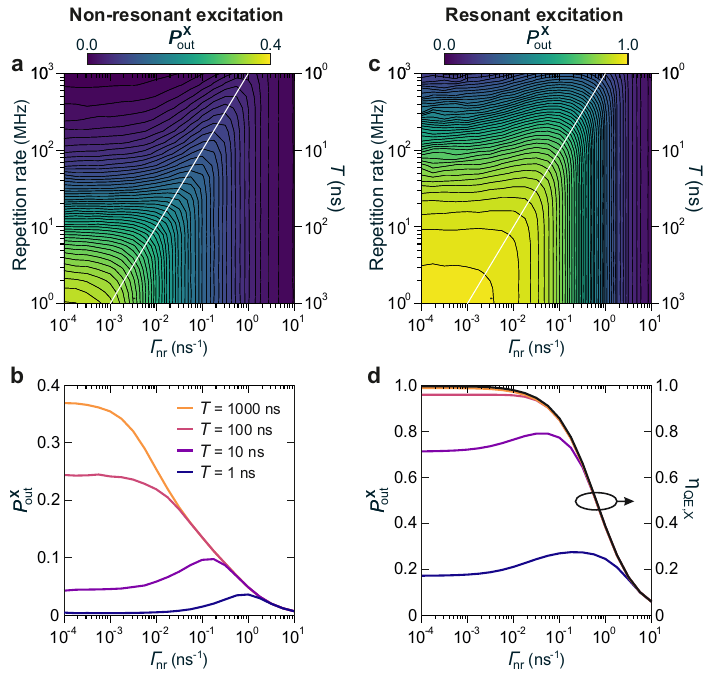} 
    \caption{Effect of the repetition rate on the probability of emission from excitons in non-resonant and resonant excitation. \textbf{a},\textbf{c} Probability of emission as a function of $\Gamma_\textrm{nr}$ and repetition rate (left vertical axis) in the non-resonant and resonant excitation. The right vertical axis shows the corresponding repetition periods ($T$). In the non-resonant regime, the QD is excited at saturation, $P_{\textrm{in}}=1.5$. The white lines indicate where $\Gamma_\textrm{nr}=1/T$. \textbf{b},\textbf{d} Probability of emission (left axis) for four different values of $T$ and internal quantum efficiency, $\eta_\text{QE,X}$, (right axis) as a function of $\Gamma_\textrm{nr}$. Note the different color scales and vertical axes.}
\label{Fig:5}
\end{figure}

The striking deviation of the emission probability in Figure~\ref{Fig:4} from the internal quantum efficiency is due to incoherent trapping of the carriers in metastable dark states, which gives rise to blinking~\cite{efros2016origin}. Blinking is not captured by a direct solution of Eq.~(\ref{eq1}), but can be effectively modeled with our stochastic solver. Figure~\ref{Fig:3}a-d shows the autocorrelation function, $g^{(2)}(\tau)$, calculated in the resonant-excitation regime for $\Gamma_\textrm{sf}=0.01~\text{ns}^{-1}$ and different values of $\Gamma_\textrm{nr}$. We observe bunching at $\tau=\pm T$ for $\Gamma_\textrm{nr}=0.1$~ns$^{-1}$ and, for even lower $\Gamma_\textrm{nr}$, an exponentially decaying autocorrelation. Further details on the long-timescale behaviour of $g^{(2)}(\tau)$ are included in the Supplementary Information.

To delve deeper into the physics and the timescales of the incoherent population trapping, we calculate histograms of the blinking duration as displayed in Figs.~\ref{Fig:4}e-h. Notably, although g$^{(2)}(\tau)$ in Fig.~\ref{Fig:4}a shows no bunching at $\tau = \pm T$, which is often interpreted as the absence of blinking, the corresponding histogram (Fig.~\ref{Fig:4}e) reveals significant uncorrelated blinking.
Furthermore, Fig.~\ref{Fig:4}f shows less blinking as compared to Fig.~\ref{Fig:4}e although the peaks in the corresponding g$^{(2)}(\tau)$ (Fig.~\ref{Fig:4}b) are not of equal height. This implies that bunching in the g$^{(2)}$-function should not always be associated with increased blinking. Interestingly, the number of blinking events decreases single-exponentially for long $\Gamma_{\textrm{nr}}>0.01$~ns$^{-1}$ (Figs.~\ref{Fig:4}e-g) and double-exponentially at $\Gamma_{\textrm{nr}}<0.01$~ns$^{-1}$ (Fig.~\ref{Fig:4}h), see corresponding fits in Supplementary Information. This observation reveals two blinking mechanisms: Bright excitons turning dark via spin-flips and QD ionization via non-radiative carrier decay. We also conclude that an increased $\Gamma_\textrm{nr}$ can neutralize the emitter by removing trapped carriers and, therefore, reduce the blinking duration. While the overall number of blinking events is generally smaller for smaller $\Gamma_\textrm{nr}$, the blinking duration can in fact be longer, which results in the population being trapped in a dark state for longer times.

Finally, we calculate $P_\textrm{out}^\textbf{X}$ as a function of the repetition rate, $1/T$, and $\Gamma_{\textrm{nr}}$ for $\Gamma_{\textrm{sf}}=0.01~\text{ns}^{-1}$. The results for both non-resonant and resonant excitation are summarized in Fig.~\ref{Fig:5}. We observe a pronounced resonant feature for non-resonant excitation when $\Gamma_\textrm{nr}T\approx 1$ (see Figs~\ref{Fig:5}a-b) and a similar resonance under resonant excitation, which gradually vanishes when $T\ge100~\text{ns}^{-1}$ (Figs \ref{Fig:5}c-d). For resonant excitation, we interpret this resonance as the result of a trade-off between the two effects: First, a reduction of the quantum efficiency, which dominates for $\Gamma_\textrm{nr} \gg 1/T$ and reduces the emission probability, and second, a suppression of the blinking duration by sweeping out carriers, which dominates for $\Gamma_\textrm{nr} \ll 1/T$. The highest emission probability occurs when $\Gamma_\textrm{nr} \sim 1/T$.
This implies a significant experimental bias in the vast majority of previous experiments on single QDs: Experimentalists tend to pick the brightest exciton lines under continuous-wave above-band excitation, but our model shows that this procedure selects the quantum dots with the lowest quantum efficiency. 

\section{Discussion}\label{sec:conc}

Our work shows that not only external defects~\cite{davancco2014multiple,efros2016origin} but also
incoherent population trapping due to carrier dynamics intrinsic to a quantum emitter can be a source of very significant blinking. This implies that the brightest emitters are not those with the highest quantum efficiency. For pulsed excitation, the repetition rate is the main factor determining which emitters will be brightest. For example, for realistic radiative and spin-flip rates of 1~ns$^{-1}$ and  0.01~ns$^{-1}$, respectively, and a repetition rate of 100~MHz, we find that the non-radiative rate leading to the brightest emission is 0.4~ns$^{-1}$ (extracted from Fig.~\ref{Fig:5}d).

Although our findings may at first glance seem surprising, there is mounting experimental evidence supporting them. First, consider time-resolved ensemble measurements on QDs. It is well-known that the non-radiative and spin-flip rates vary considerably between QDs, so the decay curve from an ensemble should be multi-exponential, but they are fitted very well with bi-exponential models~\cite{stobbe2009frequency}. Our results suggest that this is simply because the chosen repetition rate makes a particular subset of emitters much brighter. As a second comparison to experiments, we consider refs.~\cite{stobbe2009frequency} and \cite{johansen2010probing}, which both concerned the same data set: The repetition rate was 80~MHz, the measured radiative decay rate was 0.9--1~ns$^{-1}$, and the spin-flip rate was 0.006--0.015~ns$^{-1}$. For these parameters, our model predicts that the brightest QDs correspond to a non-radiative rate of 0.1--0.16~ns$^{-1}$, which closely matches the experimentally obtained value of 0.11$\pm$0.03~ns$^{-1}$. Notably, this agreement is obtained without any free parameters, which shows the remarkable predictive power of our model. Another prediction of our model is that increasing the repetition period should increase the brightness of the exciton subensensemble with a slower non-radiative decay rate, resulting in a slower measured non-radiative decay rate. This has indeed been observed but was previously unexplained~\cite{johansen2008decay}. Yet another remarkable consequence of our model is that continuous-wave excitation, which may be interpreted as the limit of the repetition period going to zero in Fig.~\ref{Fig:5}a-b, enhances the emission from the QDs with the lowest quantum efficiency. This has also been found in experiments: In Ref.~\cite{pregnolato2019deterministic}, a number of QDs that were particularly bright under continuous-wave above-band excitation were identified, but their quantum efficiencies turned out to be much lower than that of ensembles. Our work, therefore, shows that previous attempts at deterministic fabrication of photonic nanostructures around QDs, unfortunately, selected the worst QDs by using continuous-wave excitation. As a result, they are far from deterministic under resonant excitation. We can now devise a greatly improved strategy: The best QDs are the brightest under pulsed excitation with very slow repetition rates. A related approach would be to select quantum dots for which the spectrum integrated across all excitonic complexes is the brightest. As an alternative, one could use a scanning cavity with resonant pulsed excitation, as this also selects QDs with a high quantum efficiency. We note that Ref.~\cite{antoniadis2023cavity} reported on exactly such an experiment, and our model indicates that it is hardly a coincidence that this resulted in the brightest single-photon so far demonstrated. Since nearly all previous experiments on QDs unintentionally selected the worst QDs, it is likely that much brighter QDs are already present in existing devices.

Finally, while the full range of optical parameters of the QLS used in our study has not been experimentally observed in self-assembled QDs, it is relevant to other quantum emitter platforms, such as colloidal QDs~\cite{knowles2011multi,melnychuk2021multicarrier,califano2015origins}, organic fluorophores~\cite{green2015solid,nau2002biomolecular,berezin2010fluorescence}, and color centers, including those in carbon~\cite{inam2013emission,ulbricht2018excited} and silicon~\cite{baron2022detection,andrini2024activation,jhuria2024programmable,johnston2024cavity}, which hold strong potential for scalable single-photon sources compatible with silicon photonics. We anticipate that extending our stochastic model to these emitters may lead to new insights and more generally greatly accelerate the development of new quantum light sources. 

\section{Author contributions}\label{sec:contri}
S.L. wrote the code, carried out the numerical calculations, and prepared the figures. S.S. initiated and supervised the project and introduced the first version of the stochastic solver. S.L. and S.S. jointly developed the theoretical concepts and wrote the manuscript.

\section{Acknowledgements}
We express our gratitude to Niels Gregersen, Battulga Munkhbat, and Jesper Mørk for useful discussions. We gratefully acknowledge financial support from the European Research Council (Grant No.\ 101045396 -- SPOTLIGHT), the Danish National Research Foundation (Grant No.\ DNRF147 -- NanoPhoton), and the
Innovation Fund Denmark (Grant No.\ 4356-00007B -- EQUAL)

\bibliography{MAIN_no_supplementaty_codes}

\end{document}


\preprint{APS/123-QED}

\title{Supplementary information: Incoherent population trapping in quantum emitters}

\author{Sergei Lepeshov}
\affiliation{DTU Electro, Department of Electrical and Photonic Engineering, Technical University of Denmark, Ørsteds Plads 343, DK-2800, Kgs.\ Lyngby, Denmark}
\email{serle@dtu.dk}

\author{Søren Stobbe}
\affiliation{DTU Electro, Department of Electrical and Photonic Engineering, Technical University of Denmark, Ørsteds Plads 343, DK-2800, Kgs.\ Lyngby, Denmark}
\affiliation{NanoPhoton — Center for Nanophotonics, Technical University of Denmark, Ørsteds Plads 345A, DK-2800, Kgs.\ Lyngby, Denmark}
\email{ssto@dtu.dk}

\date{\today}

\maketitle
\section{\label{sec:S12}Blinking histograms in a semi-logarithmic scale}
\FloatBarrier
\renewcommand{\thefigure}{S\arabic{figure}}
\begin{figure}[H]
    \includegraphics[scale=0.88]{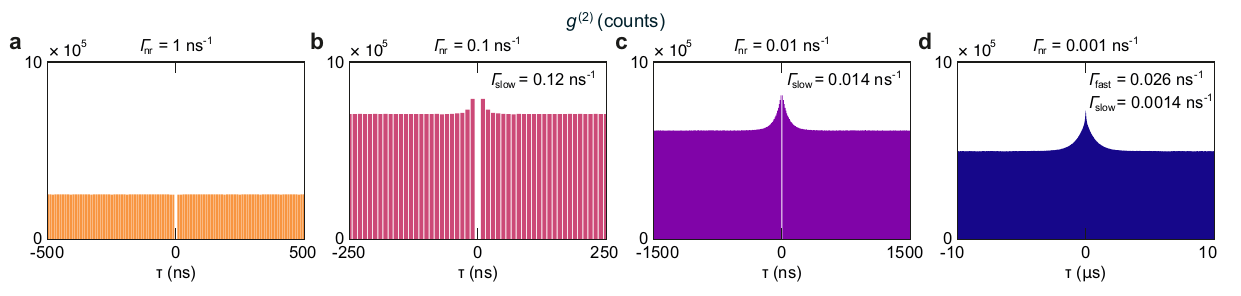}
    \caption{Long-timescale $g^{(2)}$-function for different $\Gamma_\textrm{nr}$: \textbf{a} $\Gamma_\textrm{nr}=1~\text{ns}^{-1}$, \textbf{b} $\Gamma_\textrm{nr}=0.1~\text{ns}^{-1}$, \textbf{c} $\Gamma_\textrm{nr}=0.01~\text{ns}^{-1}$, and \textbf{d} $\Gamma_\textrm{nr}=0.001~\text{ns}^{-1}$. The values of $\Gamma_{\textrm{fast}}$ and $\Gamma_{\textrm{slow}}$ indicated in each panel are extracted from single- and double-exponential fitting. The presented data are acquired for $\Gamma_\textrm{r}=1~\text{ns}^{-1}$ and $\Gamma_\textrm{sf}=0.01~\text{ns}^{-1}$}
    \label{Fig:S1}
\end{figure}

\section{\label{sec:S12}Blinking histograms in semi-logarithmic scale}
\FloatBarrier
\renewcommand{\thefigure}{S\arabic{figure}}
\begin{figure}[H]
    \includegraphics[scale=0.88]{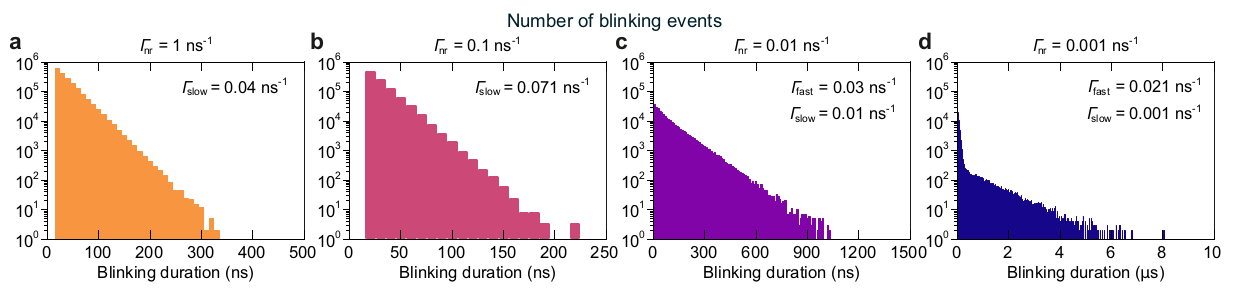}
    \caption{Blinking-duration bar charts as a function of $\Gamma_\textrm{nr}$: \textbf{a} $\Gamma_\textrm{nr}=1~\text{ns}^{-1}$, \textbf{b} $\Gamma_\textrm{nr}=0.1~\text{ns}^{-1}$, \textbf{c} $\Gamma_\textrm{nr}=0.01~\text{ns}^{-1}$, and \textbf{d} $\Gamma_\textrm{nr}=0.001~\text{ns}^{-1}$ in linear scales. The values of $\Gamma_{\textrm{fast}}$ and $\Gamma_{\textrm{slow}}$ indicated in each panel are extracted from single- and double-exponential fitting of the bar charts. The presented data are acquired for $\Gamma_\textrm{r}=1~\text{ns}^{-1}$ and $\Gamma_\textrm{sf}=0.01~\text{ns}^{-1}$}
    \label{Fig:S2}
\end{figure}

\bibliography{Bibliography}